\documentstyle[psfig]{mn}

\newif\ifAMStwofonts

\ifoldfss
  \ifCUPmtlplainloaded \else
    \NewTextAlphabet{textbfit} {cmbxti10} {}
    \NewTextAlphabet{textbfss} {cmssbx10} {}
    \NewMathAlphabet{mathbfit} {cmbxti10} {} 
    \NewMathAlphabet{mathbfss} {cmssbx10} {} 
  \fi
  \ifAMStwofonts
    \ifCUPmtlplainloaded \else
      \NewSymbolFont{upmath} {eurm10}
      \NewSymbolFont{AMSa} {msam10}
      \NewMathSymbol{\upi}     {0}{upmath}{19}
      \NewMathSymbol{\umu}     {0}{upmath}{16}
      \NewMathSymbol{\upartial}{0}{upmath}{40}
      \NewMathSymbol{\leqslant}{3}{AMSa}{36}
      \NewMathSymbol{\geqslant}{3}{AMSa}{3E}

    \fi
  \fi
\fi 

\ifnfssone
  \newmathalphabet{\mathit}
  \addtoversion{normal}{\mathit}{cmr}{m}{it}
  \addtoversion{bold}{\mathit}{cmr}{bx}{it}
  \newmathalphabet{\mathbfit} 
  \addtoversion{normal}{\mathbfit}{cmr}{bx}{it}
  \addtoversion{bold}{\mathbfit}{cmr}{bx}{it}
  \newmathalphabet{\mathbfss} 
  \addtoversion{normal}{\mathbfss}{cmss}{bx}{n}
  \addtoversion{bold}{\mathbfss}{cmss}{bx}{n}
  \ifAMStwofonts
    \ifCUPmtlplainloaded \else
      \UseAMStwoboldmath
      \makeatletter
      \new@mathgroup\upmath@group
      \define@mathgroup\mv@normal\upmath@group{eur}{m}{n}
      \define@mathgroup\mv@bold\upmath@group{eur}{b}{n}
      \edef\UPM{\hexnumber\upmath@group}
      \new@mathgroup\amsa@group
      \define@mathgroup\mv@normal\amsa@group{msa}{m}{n}
      \define@mathgroup\mv@bold\amsa@group{msa}{m}{n}
      \edef\AMSa{\hexnumber\amsa@group}
      \makeatother
      \mathchardef\upi="0\UPM19
      \mathchardef\umu="0\UPM16
      \mathchardef\upartial="0\UPM40
      \mathchardef\leqslant="3\AMSa36
      \mathchardef\geqslant="3\AMSa3E
    \fi
  \fi
\fi 

\ifnfsstwo
  \DeclareMathAlphabet{\mathbfit}{OT1}{cmr}{bx}{it}
  \SetMathAlphabet\mathbfit{bold}{OT1}{cmr}{bx}{it}
  \DeclareMathAlphabet{\mathbfss}{OT1}{cmss}{bx}{n}
  \SetMathAlphabet\mathbfss{bold}{OT1}{cmss}{bx}{n}
  \ifAMStwofonts
    \ifCUPmtlplainloaded \else
      \DeclareSymbolFont{UPM}{U}{eur}{m}{n}
      \SetSymbolFont{UPM}{bold}{U}{eur}{b}{n}
      \DeclareSymbolFont{AMSa}{U}{msa}{m}{n}
      \DeclareMathSymbol{\upi}{0}{UPM}{"19}
      \DeclareMathSymbol{\umu}{0}{UPM}{"16}
      \DeclareMathSymbol{\upartial}{0}{UPM}{"40}
      \DeclareMathSymbol{\leqslant}{3}{AMSa}{"36}
      \DeclareMathSymbol{\geqslant}{3}{AMSa}{"3E}
    \fi
  \fi
\fi 

\ifCUPmtlplainloaded \else
  \ifAMStwofonts \else 
    \def\upi{\pi}
    \def\umu{\mu}
    \def\upartial{\partial}
  \fi
\fi

\title{Intranight optical variability of radio-quiet and radio lobe dominated quasars}
\author[Stalin et al. ]
       {C.~S.~Stalin$^{1,2}$\footnote{e-mail: stalin@ncra.tifr.res.in},
 Gopal-Krishna$^1$, Ram~Sagar$^2$, and  Paul~J.~Wiita$^3$ \\
        $^1$National Centre for Radio Astrophysics, TIFR, Pune
                University Campus, 
Post Bag No.\ 3, Ganeshkhind,   Pune 411007, India
\\
        $^2$State Observatory, Naini Tal, 263129,
            India  \\
        $^3$Department of Physics \& Astronomy, MSC 8R0314, Georgia State
                University, 
        Atlanta, Georgia 30303-3088, USA
        }
\date{Accepted 2003 March xxx.
      Received 2003 March xxxx
      }

\pagerange{\pageref{firstpage}--\pageref{lastpage}}
\pubyear{2003}

\begin{document}

\maketitle

\label{firstpage}

\begin{abstract}

We present results of a programme of multi-epoch, intra-night optical
monitoring of a sample of non-blazar type AGN, which includes seven radio-quiet 
QSOs (RQQs) and an equal number of radio-loud, lobe-dominated quasars (LDQs),
covering a redshift range from about 0.2 to 2.0. These two sets of optically 
bright and intrinsically luminous QSOs are well matched in the redshift--optical 
luminosity ($z - M_B$) plane. Our CCD monitoring covered a total of 61 nights
 with an average of 6.1 hours of densely sampled monitoring of just a single 
QSO per night, thereby achieving a typical detection threshold of $\sim 1$\% 
variation over the night. Unambiguous detection of intra-night variability (INOV) 
amplitudes in the range 1$-$3\% on day-like or shorter time scales were thus 
made for both RQQs and LDQs.  Based on these clear detections of INOV, we 
estimate duty cycles of 17\% and 9\% for RQQs and LDQs, respectively; inclusion
of the two cases of probable variations of LDQs would raise the duty cycle to 
15\% for LDQs. The similarity in the duty cycle and amplitude of INOV for the
RQQs and LDQs suggests, firstly, that 
the radio loudness alone does not guarantee an enhanced INOV in QSOs and,
secondly, that as in LDQs, relativistic jets may also be present in RQQs. We argue
that, as compared to BL Lacs, the conspicuously milder, rarer and possibly slower 
INOV of RQQs and LDQs, can in fact be readily understood in terms of 
their having optical synchrotron jets which are modestly misaligned from us, 
but are otherwise intrinsically as relativistic and active as the jets in BL Lacs. 
This points toward an orientation-based unifying scheme for 
the INOV of radio-loud and radio-quiet quasars. Variability of up to 
$\sim 0.3$-mag on month to year-like time scales is seen for nearly all those 
RQQs and LDQs in our sample for which sufficient temporal coverage is available. 
These data have revealed an interesting event that seems most likely explained 
as an occultation,  lasting less than six months, of much of the nuclear optical 
continuum source in an RQQ. The observations reported here form part of a larger 
ongoing project to study the intra-night optical variability of four major 
classes of powerful AGN, including blazars.

\end{abstract}


\section{Introduction}

Multi-wavelength studies of intensity variations of quasars have played a key 
role in probing the physical conditions near the centres of activity in the 
nuclei of galaxies and in placing powerful constraints on their models, 
especially when intra-night timescales are probed. Optical variability on 
hour-like timescales for blazars has been a well established phenomenon for 
over a decade (Miller, Carini \& Goodrich 1989; Carini et al.\  1991), though 
its origin and relation to longer term variability remains unclear (e.g.\ Wiita 
1996).  A related outstanding question is the dichotomy between radio-loud 
quasars (RLQs) and radio-quiet quasars (RQQs).
In the jet dominated subset of RLQs, usually denoted as blazars, variability 
is strong in essentially all electromagnetic bands, and is commonly associated 
with the non-thermal Doppler boosted emission from jets (e.g.\ Blandford \& 
Rees 1978; Marscher \& Gear 1985; Hughes, Aller \& Aller 1990; Wagner \& 
Witzel 1995).  Intranight variability in blazars may well arise from 
instabilities or fluctuations in the flow of such jets (e.g.\ Hughes, Aller 
\& Aller 1990; Marscher, Gear \& Travis 1992).

As for  RQQs, which follow the radio-far IR correlation defined for disk 
galaxies, it has been argued that starbursts make the dominant contribution 
to the radio output in these objects (Sopp \& Alexander 1991;
Terlevich et al.\ 1992; also see Antonucci,
Barvainis \& Alloin 1990).  This finds support from the observed 
strong correlation between $\gamma$-ray detection (using EGRET) and radio 
loudness (e.g.\ Bregman 1994; see Gopal-Krishna, Sagar \& Wiita 1995).
In this
case, accretion disk instabilities may be responsible for any rapid
fluctuations detected in RQQs (e.g. Zhang \& Bao 1991; Mangalam \& Wiita 1993;
Kawaguchi et al.\ 1998).
On the other hand, jet-like radio features, or faint radio structures,
which in some cases extend far beyond the confines of the parent
galaxy, have been detected in deep radio images of several RQQs, arguing
for the existence of weak jets even in RQQs (e.g.\ Kellermann et al.\
1994; Miller, Rawlings \& Saunders 1993; Papadopulos et al.\ 1995; Kukula 
et al.\ 1998; Blundell \& Beasley 1998; Blundell \& Rawlings 2001).  The 
existence of incipient nuclear jets in RQQs has also been inferred by 
Falcke, Patnaik \& Sherwood (1996), from radio spectral measurements of 
optically selected quasar samples. The recent clear detection of intra-night 
optical variability (INOV) of a few {\it bona-fide} RQQs is readily explained 
in terms of relativistic jets on the optically emitting length scales 
(Gopal-Krishna et al.\ 2003; hereafter GSSW03). 

Two of the much debated and intriguing questions concerning AGN are the 
reality and origin of the apparent dichotomy in radio emission of  QSOs. 
Although it has long been claimed that radio-loud quasars are only a small 
fraction (10--15\%) of all QSOs, an analysis of the FIRST radio survey 
results (White et al. 2001) argued that the claimed dichotomy was an artefact 
of selection effects and that there was a continuous distribution in radio 
loudness. A similar conclusion has been reached in a recent analysis of the
radio properties of the 2dF QSO redshift survey (Cirasuolo et al.\ 2003),
employing the FIRST radio survey (Becker, White \& Helfand\ 1995). 
In contrast, another recent study of the correlations between the FIRST 
radio and preliminary SDSS optical surveys found that the dichotomy is real 
and that radio-loud sources are about 8\% of the total QSO population 
(Ivezi{\'c} et al.\ 2002). While the observational situation remains confused,
 a large number of models have been put forward to explain the RL/RQ dichotomy,
some of them based on the idea that more rapidly spinning black holes produce 
powerful relativistic jets (e.g.\ Blandford 2000; Wilson \& Colbert 1995). 
Others models argue that the radio emission correlates with the mass of the
nuclear black hole (e.g.\ Dunlop et al.\ 2003); however, this 
assertion has been questioned (Ho 2002; Woo \& Urry 2002).  
Yet other models stress 
the importance of accretion rate and possible changes in accretion mode to this 
dichotomy (e.g.\ McLure \& Dunlop 2001).

We have been pursuing the question INOV in RQQs  for  a decade now, 
expecting that any intrinsic differences between the central engines of RL 
and RQ classes of AGN could be reflected in their short-term optical 
variability.  We have carried out CCD monitoring of over a dozen optically 
luminous, bright RQQs, beginning with the first such attempt to do so
reported in Gopal-Krishna, Wiita \& Altieri (1993).  Our subsequent 
work used the 2.34m 
Vainu Bappu telescope of the Indian Institute of Astrophysics, Bangalore; 
later the 1.04m Sampurnanand telescope of the State Observatory, Naini Tal, 
was increasingly employed (Gopal-Krishna, Sagar \& Wiita 1993, 1995, 
hereafter Papers I and II, respectively; Sagar, Gopal-Krishna \& Wiita 1996, 
Paper III; Gopal-Krishna et al.\ 2000, Paper IV).  While we found several 
reasonably persuasive incidences of INOV for some RQQs, it was clearly 
necessary to extend this study through a more sensitive and systematic 
programme to confirm and better characterize this phenomenon.

This endeavour involved an optical monitoring campaign lasting 
113 nights from November 1998 through May 2002 of a matched sample 
(in both apparent magnitude and redshift) of radio-quiet and several classes
 of radio-loud quasars (BL Lacertae objects, BL; core dominated quasars,
CDQs; and  lobe dominated
quasars, LDQs).  Here we 
present our key results on the nature of INOV for the non-blazar subset 
comprised of seven RQQs and seven LDQs. A brief report on some of these 
results has been published recently (GSSW03). In a forthcoming paper we 
shall be reporting results 
for the BL Lac and CDQ components of this program (Sagar et al.\ 2003). 
Additional details can be found in Stalin (2002) and Stalin et al (2003).

\section[]{Current status of INOV in RQQs}

In Papers III and IV, where a fairly dense temporal sampling was achieved,
we reported several instances of apparently significant INOV for RQQs.  
These events 
could be classified as small gradual variations lasting over several hours
(e.g. 1630$+$377), time resolved microvariability on hour-like timescales 
(e.g. 0946$+$301, 1444$+$407), and single-point fluctuations, designated as 
``spikes'' (e.g. \ 0748$+$294, 0824$+$098, 1444$+$407, 1630$+$377).

Other groups also have made attempts to detect and characterize INOV among radio 
quiet AGN (Petrucci et al.\ 1999; Rabbette et al.\ 1998) and among
mixed samples of RQ and RL AGN (Jang \& Miller 1995, 1997; de Diego et al.\ 
1998; Rabbette et al.\ 1998; Romero, Cellone \& Combi 1999). Even if the
variations claimed in these works were correctly identified, it appeared that
both the INOV amplitude and the duty cycle of the RQQs were small compared 
to those found in many studies of  blazars (e.g.\ Carini et al. 1992; Heidt 
\& Wagner 1998; Xie et al.\ 2002; Romero et al.\ 2002).

Among the other studies of RQ AGN variability, that of Petrucci et al.\ 
(1999) exclusively involved Seyfert 1 galaxies, which are much weaker 
than the sources we consider here. Because the AGN contribution to the total 
light is minor, seeing  variations can greatly complicate
accurate detections of variations in the nuclei of Seyferts. The programme of
Rabbette et al.\ (1998) involved BVR monitoring of 23 high luminosity RQQs, 
22 of which are at $z > 1$. While the basic approach of using 2--3 comparison 
stars within the CCD frame, as well as keeping the exposure time at around 
10 minutes, are similar to the present work, there are major differences 
as well. Firstly, in the program of Rabbette et al. (1998), 
intra-night sampling was usually much sparser, with only a few data points 
per object per  night. This, coupled with their typical rms noise of 4\% 
raises their microvariability detection threshold to $\sim$0.1 mag, which 
is many times higher than that attained in our observations. The same large 
errors also hamper their attempts to detect longer-term variability. We believe 
that this factor can explain their total lack of detection of intra-night 
variations and the near absence of night-to-night or longer-term variability 
in their observations of RQQs. Thus, given the differences in their 
observational approach and instrumental sensitivity, the results of the two
campaigns are not discrepant.

Over the past several years a number of independent studies have been carried 
out to investigate the difference between the INOV in RL and RQ AGNs, with the 
goal of constraining models of the RL/RQ dichotomy. Jang \& Miller (1995, 1997) 
studied a total sample of 19 RQ AGN and 11 RL AGN and found INOV in 3 (16\%) 
of the former and 9 (82\%) of the latter. However, optical luminosities of 
these RQ AGNs are modest, M$_B$ $>$ $-$24.3, and close to the critical value 
below which the radio properties are thought to become like those of Seyfert 
galaxies (Miller, Peacock \& Mead 1990); hence they are not the 
{\it bona-fide} quasars which are our primary concern here.

Romero et al.\ (1999) monitored a sample of 23 southern quasars: 8 RQQs and 
15 blazars.  The details of the production of their differential light curves
differed somewhat from those of Papers I--IV and of Jang \& Miller (1995,
 1997), particularly in their averaging of 6 comparison stars to produce two 
effective comparison objects.  Still, this approach should provide basically 
very similar results unless one or more of their comparison stars also showed 
substantial INOV, in which case their stellar errors will be too large and 
their detection threshold for AGN variability will be too high.  None of their
8 RQQs was found to vary down to 1\% rms, while 9 of the 15 blazars showed INOV.  
Romero et al.\ (1999) enlarged their above-mentioned sample by including
the objects monitored by us in Paper II and by Jang \& Miller (1995, 1997).
This enlarged sample contained 27 RQQs and 26 RLQs and they derived duty 
cycles for the RLQs and RQQs of above 70\% and only 3\%, respectively, from 
this mixed sample.  Here ``duty-cycle'' is defined as the ratio of the 
observational sessions during which objects of the particular class are 
detected as variable to the total observing time spent on objects in that 
class.

In contrast to the results summarized so far, de Diego et al.\ (1998) 
concluded that microvariability is as at least as common among RQQs as it 
is among the (relativistically beamed) CDQ sources, commonly deemed as 
blazars.  They  claimed  detections of INOV in 6 of 30 RQQ monitoring
sessions and only 5 of 30 CDQ sessions.  Their sample was chosen so that each 
of their 17 RQQs  had a  CDQ counterpart of nearly matching brightness and 
redshift.   However, their study differs radically from all other programs,
including ours, in the procedure adopted for observation and analysis.
de Diego et al.\ (1998) observed each source only between 3 and 9 times per night;
each such observation consisted of five 1-minute exposures of the target field.  
An INOV analysis was then made through an ANOVA procedure which attempts 
to determine observational errors directly from the scatter in the object 
minus a reference star for the 5 points within each set of observations.  
The other programs, on the other hand, obtained degree of significance of 
variations either from the errors given by an aperture  photometry algorithm,
after suitable calibration, or from the scatter in the differential light
curves (DLCs) of the comparison 
stars; we believe these techniques to be more reliable.

These markedly discrepant recent results prompted us to pursue this question 
by conducting an extensive programme of sensitive intra-night monitoring of 
a large sample of powerful AGN representing the four major classes mentioned 
above.

\begin{table*}
\caption{The seven sets of radio-quiet (RQQ) and lobe-dominated quasars (LDQs)
 monitored in the present programme}
\begin{tabular}{lllllllllrr} \hline 
Set No. &Object & Other Name  & Type & RA(2000)  & Dec(2000)   &  ~~B    & ~~M$_B$  & {\it ~~z} & \%Pol$^a$ & R$^b$ \\ 
        &       &             &      &           &             & (mag)   & (mag)    &           &  (optical)       &           \\ \hline
1.      &0945+438   & US 995          & RQQ  &09 48 59.4 &   +43 35 18 &  16.45  & $-$24.3  & 0.226  &  ~~~---   & $<$0.85    \\       
        &2349$-$014 & PKS 2349-01     & LDQ  &23 51 56.1 & $-$01 09 13 &  15.45  & $-$24.7  & 0.174  & 0.91      &  295       \\   
2.      &0514$-$005 & 1E 0514-0030    & RQQ  &05 16 33.5 & $-$00 27 14 &  16.26  & $-$25.1  & 0.291  & ~~~---    & $<$ 1.1    \\ 
        &1004+130   & PG  1004+130    & LDQ  &10 07 26.2 &   +12 48 56 &  15.28  & $-$25.6  & 0.240  & 0.79      &  195       \\   
3.      &1252+020   & Q  1252+0200    & RQQ  &12 55 19.7 &   +01 44 13 &  15.48  & $-$26.2  & 0.345  & ~~~---    &  0.52      \\   
        &0134+329   & 3C 48.0         & LDQ  &01 37 41.3 &   +33 09 35 &  16.62  & $-$25.2  & 0.367  & 1.41      &  8511      \\
4.      &1101+319   & TON 52          & RQQ  &11 04 07.0 &   +31 41 11 &  16.00  & $-$26.2  & 0.440  &  ~~~---   &$<-$0.39    \\ 
        &1103$-$006 & PKS 1103$-$006  & LDQ  &11 06 31.8 & $-$00 52 53 &  16.39  & $-$25.7  & 0.426  & 0.37      &  631       \\
5.      &1029+329   & CSO 50          & RQQ  &10 32 06.0 &   +32 40 21 &  16.00  & $-$26.7  & 0.560  & ~~~---    &$<-$0.23    \\
        &0709+370   & B2 0709+37      & LDQ  &07 13 09.4 &   +36 56 07 &  15.66  & $-$26.8  & 0.487  & ~~~---    & 120        \\     
6.      &0748+294   & QJ 0751+2919    & RQQ  &07 51 12.3 &   +29 19 38 &  15.00  & $-$29.0  & 0.910  & ~~~---    & 0.21       \\   
        &0350$-$073 & 3C 94           & LDQ  &03 52 30.6 & $-$07 11 02 &  16.93  & $-$27.2  & 0.962  & 1.42      & 1175       \\
7.      &1017+279   & TON 34          & RQQ  &10 19 56.6 &   +27 44 02 &  16.06  & $-$29.8  & 1.918  & ~~~---    &$<-$0.32    \\ 
        &0012+305   & B2 0012+30      & LDQ  &00 15 35.9 &   +30 52 30 &  16.30  & $-$29.1  & 1.619  & ~~~---    & 57.5       \\
\hline 
\end{tabular}

\hspace*{-10.0cm} $^a$ Reference to the polarization data: Wills et al. (1992)

\hspace*{-9.1cm} $^b$ R is the ratio of the radio-to-optical flux densities (Sect. 3.1)

\end{table*}

\section{Observations}

\subsection{The sample and instruments used} 
The  sample of non-blazar  
objects (i.e. RQQs and LDQs) considered in this paper consists of seven 
pairs of these  AGN covering a total redshift range from 0.17 to 1.92 
(Table 1), taken from the catalog of V{\'e}ron-Cetty \& V{\'e}ron (1998).  
Each pair is closely matched in both $z$ and B magnitude: their catalogued 
apparent magnitudes are $15 < m_B < 17$ and their absolute magnitudes range 
between $-24.3$ and $-29.8$ (assuming $H_0 = 50, q_0 = 0$), so all of these 
objects are {\it bona fide} QSOs. The radio properties of RQQs were 
determined from Kellermann et al.\ (1994), supplemented by the NVSS (Condon 
et al.\ 1998) and FIRST (Becker, White \& Helfand 1995) surveys at 1.4 GHz, and our 
own VLA observations of two of the RQQs at 5 GHz (1029$+$329 and 1252$+$020).
For all the seven RQQs, $R < 1$, where $R$ is the rest-frame ratio of 5 GHz to 
250 nm flux densities, computed following the prescription of Stocke et al.\
(1992).  The criteria adopted for 
an LDQ designation was a radio spectral index $\alpha < -0.5$ ($S_{\nu} 
\propto \nu^{\alpha}$) as determined either from simultaneous flux 
measurements between 1 -- 22 GHz (Kovalev et al.\ 1992) or from NED\footnote
{URL http://ned.ipac.caltech.edu/}.  For five of the seven LDQs sufficiently detailed
radio maps are available, and the core emission at 5 GHz is found to be weaker 
than the lobe emission (see Wills \& Browne 1986). Evidence for lack of strong 
beamed non-thermal emission in the optical continuum is the weak optical 
polarization ($<$ 1.5\%) known for five of the seven LDQs in our sample (Wills 
et al.\ 1992).  Additional details concerning the sample selection, including 
our VLA observations and also the results for the blazar component of our programme, 
can be found in Sagar et al.\ (2003), Stalin (2002)  and Stalin et al.\ (2003).

All observations reported here were carried out at the State Observatory,
Naini Tal, using the 104-cm  Sampurnanand telescope; this is an RC Cassegrain 
system with a f/13 beam (Sagar 1999).  The detector used for the observations
was a cryogenically cooled 2048$\times$2048 Wright CCD, except prior to 
October 1999, when a 1024$\times$1024 Tektronix CCD was in use.  In each CCD 
a pixel corresponds to $0.^{\prime\prime}38 \times 0.^{\prime\prime}38$, 
covering a $12^{\prime}\times 12^{\prime}$ field for the larger, and a $6^
{\prime}\times 6^{\prime}$ field for the smaller, CCD.  Since the CCDs' 
sensitivities peak in the R-band, a standard R filter was used for all of the 
observations, which were conducted on a total of 61 nights (29 for RQQs, 32 
for LDQs), with a typical duration of $\sim 6$ hours per night.  On each night 
only one AGN was monitored, as continuously as possible, and the typical 
sampling rate was about 5 frames per hour. The choice of exposure time depended 
on the brightness of the QSO and of the moon as well as the sky transparency.  
All these QSOs were chosen so that at least 2 -- 3 comparison stars within 
about a magnitude of the QSO 
were simultaneously registered on the CCD frame.  This redundancy allowed us 
to identify and discount any comparison star which itself varied during a 
given night, and thereby ensured reliable differential photometry of the QSO.

Table 2 gives a log of our observations of the QSOs which we found to be 
intranight variables (or, probable variables). Table 3 gives the positions
and apparent magnitudes of the comparison stars used for these and other
QSOs whose DLCs are reported in this paper. Finding charts for all the 
comparison stars used in our programme can be found in Stalin (2002) and 
Stalin et al.\ (2003).

\subsection{Data Processing}

Preliminary processing of the images, as well as the photometry, was done using
IRAF\footnote{Image Reduction and Analysis Facility, distributed by NOAO,
operated by the AURA, Inc. under  agreement with the NSF}. Bias frames
were obtained every night and the average bias frame was subtracted from all 
image frames after clipping the cosmic-ray (CR) hits. Dark frame subtraction 
was not carried out because the CCDs were cooled to $-120^\circ$C and so the
accumulation of thermal charge was negligible.  The flat-fielding of the frames 
was done by taking several twilight sky frames which were median combined to 
generate the flat-field template which was then used to derive the final frames. 
The final step involved removing CR hits seen in the flat-fielded target frames 
using the facilities available in the MIDAS\footnote{Munich Image and Data 
Analysis System, designed and developed by the ESO} software.

\begin{table}
\caption{Log of the optical observations of RQQs and LDQs}
\begin{tabular}{lllcclll} \hline
Object    & Type& Date     & N    & T   & INOV  &  $C_{\rm eff}$      & $\psi$ \\
          &     &          &      &(hr) & status&                     & (\%)   \\ \hline
0945+438  & RQQ & 15.01.99 &  44  & 8.0 &  NV   &                     &         \\
          &     & 26.02.00 &  31  & 6.3 &  NV   &                     &        \\
          &     & 23.01.01 &  24  & 6.6 &  NV   &                     &        \\
2349-014  & LDQ & 13.10.01 &  34  & 6.8 &  V    &  3.6                & 2.2    \\
          &     & 17.10.01 &  39  & 7.6 &  V    &  3.1                & 1.5    \\
          &     & 18.10.01 &  40  & 7.7 &  V    &  3.2                & 1.6    \\
0514$-$005& RQQ & 09.12.01 &  25  & 5.3 & NV    &                     &        \\
          &     & 10.12.01 &  23  & 5.8 & NV    &                     &        \\
          &     & 19.12.01 &  35  & 7.5 & NV    &                     &        \\
1004+130  & LDQ & 27.02.99 &  30  & 4.3 &  NV   &                     &        \\
          &     & 16.02.99 &  36  & 6.5 &  NV   &                     &        \\
          &     & 29.03.00 &  21  & 3.8 &  NV   &                     &        \\
          &     & 30.03.00 &  26  & 4.6 &  NV   &                     &        \\
          &     & 18.02.01 &  42  & 5.5 &  NV   &                     &        \\
          &     & 24.03.01 &  50  & 6.4 &  NV   &                     &        \\
 1252+020 & RQQ & 22.03.99 & 36   & 6.4 & V    &    3.3              &  2.3    \\
          &     & 09.03.00 & 29   & 6.1 & NV   &                     &        \\
          &     & 03.04.00 & 19   & 4.3 & V    &    3.6              &  0.9    \\
          &     & 26.04.01 & 20   & 4.6 & NV   &                     &        \\
          &     & 18.03.02 & 19   & 7.3 & NV   &                     &          \\
 0134+329 & LDQ & 07.11.01 & 33   & 6.5 & NV   &                     &         \\
          &     & 08.11.01 & 32   & 6.7 & NV   &                     &         \\
          &     & 13.11.01 & 46   & 8.6 & NV   &                     &         \\
 1101+319 & RQQ & 12.03.99 &  39  & 8.5 & NV   &                     &         \\
          &     & 04.04.00 &  22  & 5.6 & NV   &                     &         \\
          &     & 21.04.01 &  21  & 6.1 & V    &     2.6             & 1.2      \\
          &     & 22.04.01 &  21  & 5.8 & NV   &                     &         \\
1103$-$006& LDQ & 17.03.99 &  23  & 3.8 & NV   &                     &         \\
          &     & 18.03.99 &  40  & 7.5 & V    &    3.1              & 2.4      \\
          &     & 06.04.00 &  13  & 3.9 & PV   &    2.1              & 1.2      \\
          &     & 25.03.01 &  28  & 7.2 & NV   &                     &        \\
          &     & 14.04.01 &  19  & 4.5 & NV   &                     &         \\
          &     & 22.03.02 &  15  & 5.8 & PV   &    2.2              & 0.7        \\
 1029+329 & RQQ & 02.03.00 &  19  & 5.0 & NV   &                     &         \\
          &     & 05.04.00 &  19  & 5.3 & V    &    4.3              & 1.3      \\
          &     & 23.03.01 &  20  & 5.8 & NV   &                     &         \\
          &     & 06.03.02 &  31  & 8.5 & NV   &                     &         \\
          &     & 08.03.02 &  17  & 6.1 & V    &   2.8               & 1.1    \\
 0709+370 & LDQ & 20.01.01 &  29  & 6.5 & NV   &                     &         \\
          &     & 21.01.01 &  29  & 6.2 & NV   &                     &         \\
          &     & 25.01.01 &  31  & 7.1 & NV   &                     &         \\
          &     & 20.12.01 &  49  & 7.9 & V    &  3.1                & 1.4     \\
          &     & 21.12.01 &  48  & 7.5 & NV   &                     &         \\
 0748+294 & RQQ & 14.12.98 &  22  & 7.6 & NV   &                     &         \\
          &     & 13.01.99 &  56  & 8.3 & NV   &                     &         \\
          &     & 09.12.99 &  26  & 5.1 & NV   &                     &         \\
          &     & 24.11.00 &  28  & 5.4 & NV   &                     &         \\
          &     & 01.12.00 &  32  & 6.0 & NV   &                     &         \\
          &     & 25.12.01 &  30  & 5.4 & NV   &                     &         \\
0350$-$073& LDQ & 14.11.01 &  31  & 6.6 & NV   &                     &         \\
          &     & 15.11.01 &  26  & 5.5 & NV   &                     &         \\
          &     & 18.11.01 &  25  & 5.7 & NV   &                     &         \\
 1017+279 & RQQ & 14.03.99 &  43  & 7.3 & NV   &                     &         \\
          &     & 14.01.00 &  33  & 7.1 & NV   &                     &         \\
          &     & 27.02.00 &  33  & 8.1 & NV   &                     &         \\
 0012+305 & LDQ & 18.01.01 &  17  & 3.6 & NV   &                     &         \\
          &     & 20.01.01 &  14  & 3.2 & NV   &                     &         \\
          &     & 24.01.01 &  14  & 2.9 & NV   &                     &         \\
          &     & 14.10.01 &  20  & 5.7 & NV   &                     &         \\
          &     & 21.10.01 &  22  & 5.7 & NV   &                     &        \\
          &     & 22.10.01 &  24  & 6.2 & NV   &                     &         \\ \hline
\end{tabular}
\end{table}

\begin{table}
\caption{Position and apparent magnitudes of the comparison stars}
\begin{tabular}{llrrll} \hline
Source     &   Star  &  RA(2000)     &  Dec(2000)     &  R     &   B     \\ \hline
0945+438   &   S1    &  09 49 16.72  &    43 33 35.5  &  16.0  &  16.6   \\
RQQ(Set 1) &   S2    &  09 49 06.08  &    43 34 16.3  &  16.4  &  18.1   \\
           &   S3    &  09 49 23.91  &    43 35 02.3  &  16.9  &  18.1   \\
2349$-$014 &   S1    &  23 52 08.96  & $-$01 05 09.7  &  14 2  &  15.7   \\
LDQ(Set 1) &   S2    &  23 52 12.60  & $-$01 03 38.3  &  14.5  &  15.3   \\
           &   S3    &  23 52 13.47  & $-$01 11 07.9  &  14.7  &  16.0   \\
0514$-$005 &   S1    &  05 16 31.08  & $-$00 27 07.7  &  15.8  &  16.0   \\
RQQ(Set 2) &   S2    &  05 16 27.30  & $-$00 30 18.9  &  15.4  &  15.5   \\
           &   S3    &  05 16 44.64  & $-$00 26 46.8  &  15.7  &  16.4   \\
1004+130   &   S1    &  10 07 26.91  &    12 46 09.5  &  15.4  &  15.8   \\
LDQ(Set 2) &   S2    &  10 07 23.24  &    12 44 53.9  &  14.6  &  15.0   \\
1252+020   &   S1    &  12 55 20.98  &    01 41 13.9  &  15.2  &  15.6   \\
RQQ(Set 3) &   S2    &  12 55 35.53  &    01 41 06.7  &  15.4  &  17.1   \\
           &   S3    &  12 55 33.90  &    01 45 20.4  &  15.2  &  16.1   \\
           &   S4    &  12 55 15.61  &    01 43 54.9  &  15.0  &  15.7   \\
           &   S5    &  12 55 36.03  &    01 42 04.4  &  16.0  &  16.2   \\
           &   S6    &  12 55 33.14  &    01 45 01.6  &  15.4  &  16.8   \\
0134+329   &   S1    &  01 37 48.56  &    33 09 31.0  &  15.5  &  17.2   \\
LDQ(Set 3) &   S2    &  01 37 51.26  &    33 07 08.9  &  15.6  &  16.3   \\
           &   S3    &  01 37 37.27  &    33 03 26.5  &  15.4  &  16.1   \\
1101+319   &   S1    &  11 04 04.42  &    31 41 25.0  &  16.8  &  17.6   \\
RQQ(Set 4) &   S2    &  11 04 13.05  &    31 41 42.2  &  16.2  &  17.2   \\
           &   S3    &  11 04 10.46  &    31 43 52.8  &  16.4  &  16.8   \\
           &   S4    &  11 04 14.10  &    31 44 10.2  &  15.8  &  17.8   \\
           &   S5    &  11 04 30.14  &    31 37 20.3  &  16.5  &  18.1   \\
1103$-$006 &   S1    &  11 06 42.42  & $-$00 56 46.3  &  15.2  &  15.5   \\
LDQ(Set 4) &   S2    &  11 06 44.58  & $-$00 56 23.7  &  15.5  &  15.9   \\
           &   S3    &  11 06 32.47  & $-$00 52 41.8  &  17.1  &  19.0   \\
           &  CS3    &  11 06 29.36  & $-$00 52 44.8  &  14.0  &  15.2   \\          
1029+329   &   S1    &  10 32 08.93  &    32 37 50.7  &  15.3  &  17.4   \\
RQQ(Set 5) &   S2    &  10 31 59.48  &    32 41 56.1  &  16.3  &  16.5   \\
           &   S3    &  10 32 03.52  &    32 40 19.6  &  15.8  &  18.7   \\
           &   S4    &  10 32 07.47  &    32 37 28.1  &  15.1  &  16.3   \\
           &   S5    &  10 31 57.21  &    32 39 20.0  &  15.1  &  16.3   \\
           &   s6    &  10 32 10.74  &    32 36 06.4  &  14.8  &  15.5   \\
0709+370   &   S1    &  07 13 01.95  &    36 59 59.3  &  16.0  &  16.4   \\
LDQ(Set 5) &   S2    &  07 13 09.80  &    37 00 35.5  &  15.8  &  16.3   \\
           &  CS2    &  07 13 24.12  &    36 56 47.4  &  14.4  &  16.2   \\       
           &   S3    &  07 13 04.57  &    37 01 08.5  &  15.2  &  15.7   \\
0748+294   &   S1    &  07 50 57.78  &    29 18 20.8  &  15.9  &  16.8   \\
RQQ(Set 6) &   S2    &  07 50 56.95  &    29 17 51.5  &  15.7  &  16.4   \\
           &   S3    &  07 51 18.87  &    29 18 36.6  &  16.3  &  16.9   \\
0350$-$073 &   S1    &  03 52 39.78  & $-$07 11 12.3  &  14.6  &  15.5   \\
LDQ(Set 6) &   S2    &  03 52 40.55  & $-$07 10 10.1  &  15.3  &  15.7   \\
           &   S3    &  03 52 28.63  & $-$07 08 17.0  &  15.4  &  15.7   \\
1017+279   &   S1    &  10 19 55.67  &    27 46 09.1  &  16.1  &  18.6   \\
RQQ(Set 7) &   S2    &  10 19 42.79  &    27 44 53.2  &  15.6  &  16.3   \\
           &   S3    &  10 19 41.83  &    27 45 51.0  &  15.1  &  15.8   \\
           &   S4    &  10 19 54.59  &    27 46 14.5  &  15.0  &  15.6   \\
           &   S5    &  10 19 44.13  &    27 46 08.6  &  14.4  &  14.9   \\
0012+305   &   S1    &  00 15 38.18  &    30 52 15.6  &  16.9  &  17.8   \\
LDQ(Set 7) &   S2    &  00 15 29.85  &    30 50 38.8  &  16.6  &  17.3   \\
           &   S3    &  00 15 23.92  &    30 52 25.1  &  16.5  &  17.3   \\
           &   S4    &  00 15 39.99  &    30 50 08.2  &  15.3  &  16.4   \\
           &   S5    &  00 15 16.43  &    30 52 42.4  &  15.0  &  16.8   \\ \hline
\end{tabular}
\end{table}

On a given night, the aperture photometry of the QSO and their chosen comparison 
stars present in each frame employed the same circular aperture to determine 
instrumental magnitudes, using the {\it daofind} and {\it phot} tasks in IRAF.
The derived instrumental magnitudes were used to construct differential light 
curves (DLCs) of a given QSO relative to the chosen comparison stars as well 
as between all pairs of the comparison stars.  On each night a range of aperture 
radii were considered and the one that minimized the variance of the DLC of 
the steadiest pair of comparison stars was adopted; the mean value of the 
aperture radius used was $4.^{\prime \prime}0 \pm 1.^{\prime \prime}3$.  
We stress that the DLCs are not sensitive to the exact choice of aperture radius.

The  B and R magnitudes and colours for each QSO and its comparison star were 
obtained from the USNO 
catalog\footnote{http://archive.eso.org/skycat/servers/usnoa}; 
the difference between the $B - R$ colour indices 
of the QSO and at least one of its comparison stars was always found to be 
less than one magnitude, except for the LDQ 1103$-$006, where it was 1.2 mag 
(Table 3). 

\section{Results}

\subsection{Intra-night and inter-night variability}

In Figures 1 and 2 the derived DLCs are presented for the 3 LDQs and 3 RQQs
that showed evidence of INOV on 12 nights. Out
of these, 10 DLCs show clear INOV, while the remaining two DLCs are 
classified as `Probable Variable (PV)' (Table 2). All these 12 DLCs display
3-point running averages of the respective original data sequences, taken for 
improving the 
signal/noise ratio. However, no averaging of the data points was done for 
the DLCs meant to show the observed single-point spikes which are displayed
in Fig.\ 3.
The entire set of DLCs obtained in our programme (113 nights) is presented 
in Stalin (2002) and Stalin et al.\ (2003).  We note that the error
bars shown on the data points are those given by the {\it phot} algorithm in
IRAF; however, these nominal error bars are, for DAOPHOT reductions,
too small by a factor of $\sim$ 1.5 (Stalin 2002; GSSW03). So 
we have used this correction factor in the further analysis.  

\begin{figure*}
\vspace*{-7.0cm}
\hspace*{-0.5cm}\psfig{file=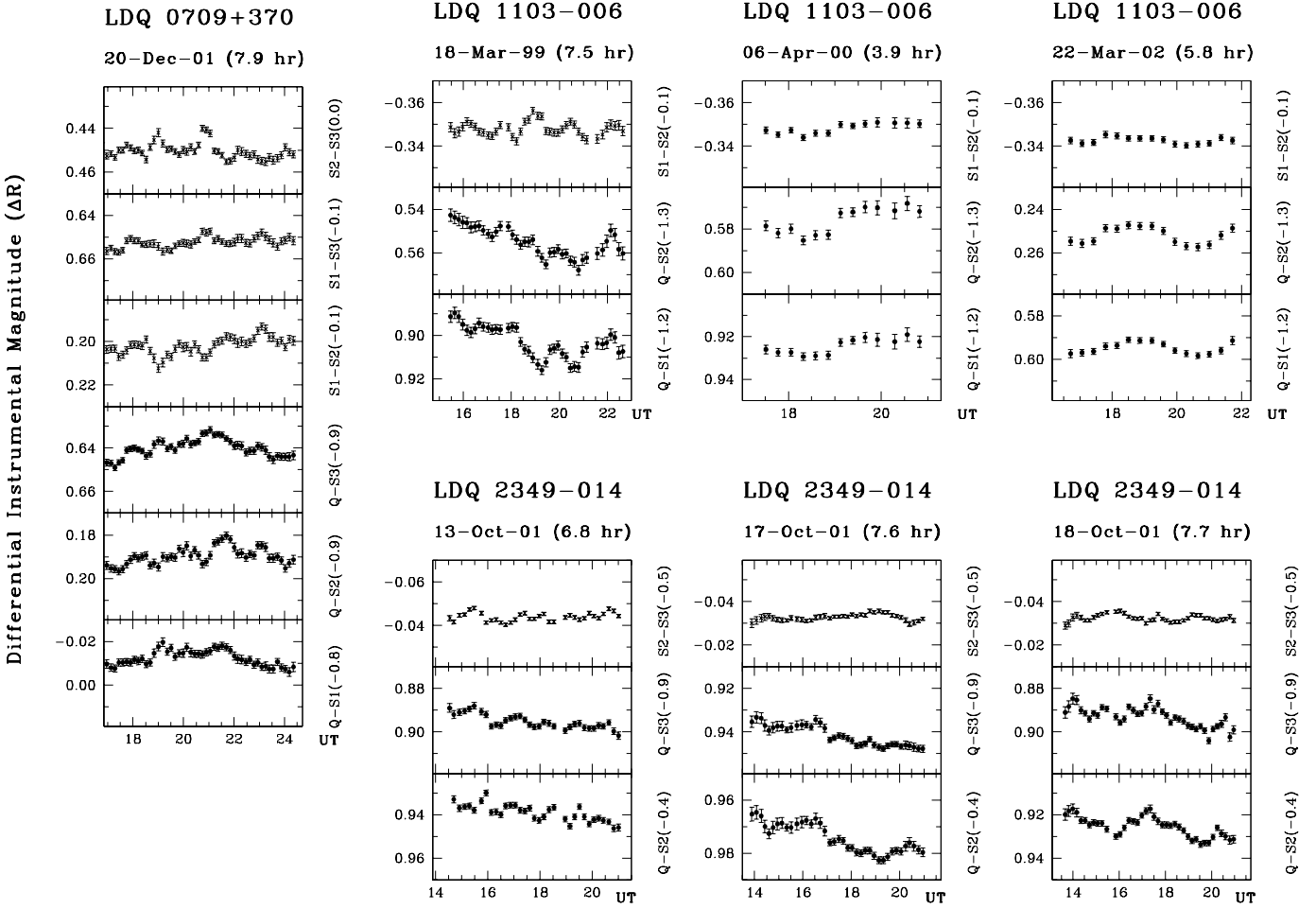,width=19.0cm}
\vspace*{-6.0cm}
\caption{
 Differential light curves (DLCs) for the radio lobe-dominated quasars (LDQs)
with a positive or probable detection of INOV.
The name of the quasar, the date and the duration of observations are given
at the top of each night's observations. The upper panel(s) give the 
differential light curves (DLCs) for the various pairs of comparison stars
available and the subsequent panels give the quasar-star DLCs, as defined
in the labels on the right side. The numbers inside the parentheses are the 
differential color indices, $\Delta (B-R)$ for the respective DLCs.
}
\end{figure*}

\begin{figure*}
\vspace*{-3.5cm}
\hspace*{-0.5cm}\psfig{file=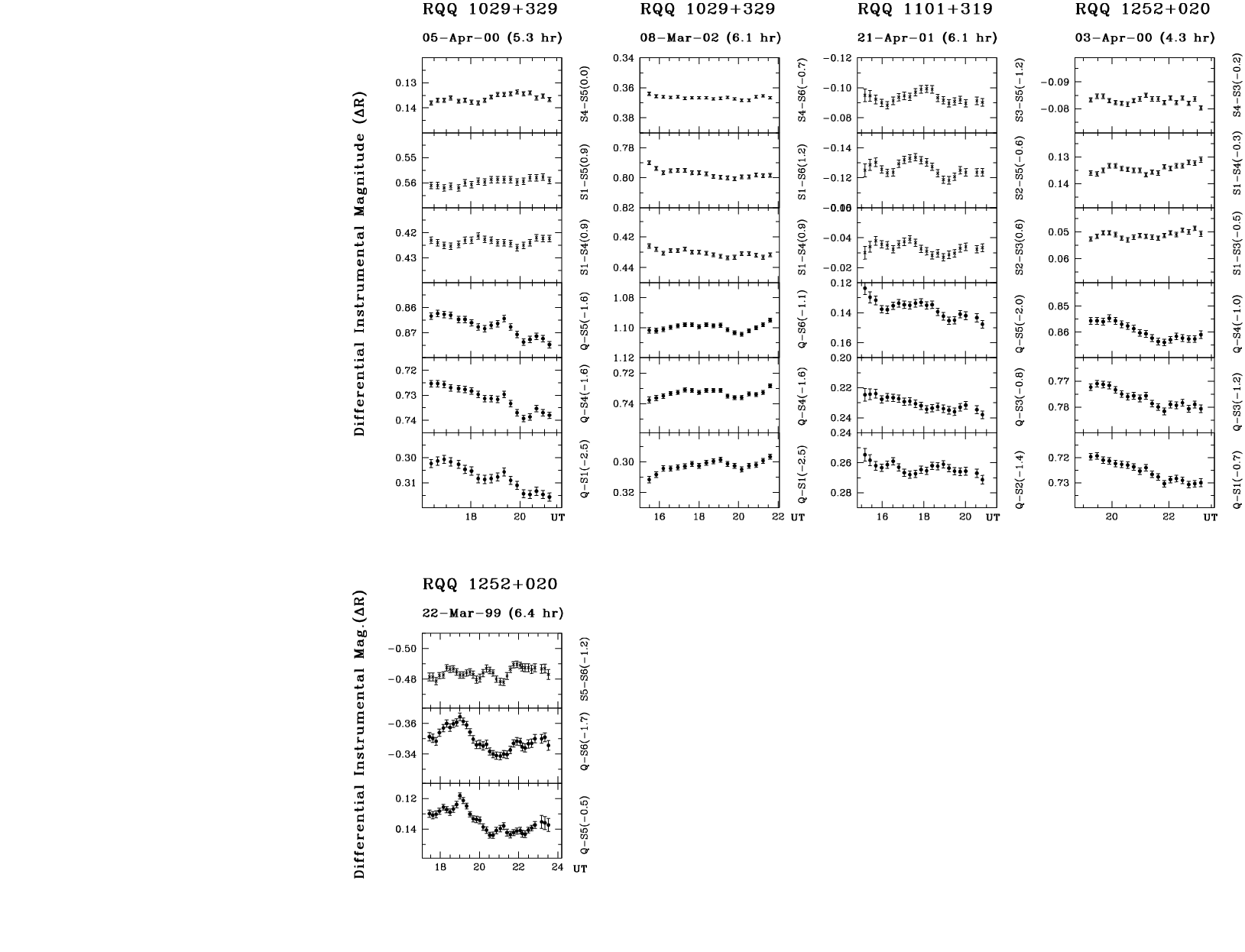,width=19.0cm}
\vspace*{-6.0cm}
\caption{
Differential light curves (DLCs) for the radio-quiet quasars (RQQs) for
which INOV was detected. The format is identical to that of Fig. 1.
}
\end{figure*}
Table 2 provides information on the variability status for each night of 
monitoring, as inferred from the DLCs. For the variable and probable variable
QSOs, we have also given the values of the parameter,
$C_{\rm eff}$ which employs an statistical criterion similar to that of Jang \&
Miller (1997) with the added advantage that for each QSO we have DLCs 
relative to multiple comparison stars (see GSSW03). This allowed us to discard
any INOV candidates for which the multiple DLCs do not show clearly
correlated trends, both in amplitude and time. For a given DLC, we take the
ratio of its standard deviation and the mean $\sigma_i$ of its individual 
data points. This ratio, $C_i$, for the $i^{th}$ DLC of a given QSO
has the corresponding probability $p_i$ that the DLC is non-variable, 
assuming a normal distribution. We then compute the joint
probability, $P$, by multiplying the values of $p_i$'s for individual DLCs
available for the QSO. This effective $C$ parameter, $C_{\rm eff}$, corresponding
to $P$, is given in Table 2 for each variable or probable variable DLC. Our
criterion for variability is $C_{\rm eff} > 2.57$, which corresponds to a
confidence level in excess of 99\%. The criterion for `probable variable'
is $C_{\rm eff} > 2.00$ which corresponds to a confidence level $>$ 95\%. 
The last column of Table 2 gives for each DLC the value of of peak-to-peak 
variability amplitude, 
$\psi \equiv [(D_{\rm max} - D_{\rm min})^2 - 2\sigma^2]^{1/2}$.
Here, $D$ is the differential magnitude, 
$\sigma^2 = \eta^2<\sigma_{\rm err}^2>$, with $\eta$ the factor by which the
 average of the measurement errors
($\sigma_{\rm err}$, as given by ${\it phot}$ algorithm)
should be multiplied; we find $\eta = 1.50$ (Stalin 2002; GSSW03).

As an added precaution, we have used the colour information on the comparison 
stars (Table 3) to check if the inferred INOV of any of the QSOs is spurious,
arising from a
combination of a large differential colour index $\Delta(B-R)$, for the QSO 
DLCs and the varying airmass with zenith distance during the night. For each 
night when the QSO DLCs showed (correlated) INOV and all of those DLCs had 
large $\Delta(B-R)$ (amplitude significantly higher than 1-mag),
the apparent INOV of the QSO could conceivably be spurious, unless at least 
one of the star-star DLC on the same night also had similarly large $\Delta(B-R)$
and yet showed no systematic trend over the night. In the present sample, a
possible such case of a QSO designated as variable is the RQQ 1029+329 
(05 April 2000). By generating a star-star DLC with a large differential colour 
index $\Delta(B-R) =$ $-$1.9 from the frames taken on the same night of
05 Apr 2000, it has already been shown in GSSW03 that even such a large value
of $\Delta(B-R)$ did not produce a systematic variation in the DLC.
Therefore, the inferred INOV of the RQQ cannot be an artefact of the similarly 
large colour differences that exist between this RQQ and its comparison stars.

Another potential source of spurious variability in such aperture photometry
is the contamination arising from the host galaxy of the target AGN.
As pointed out by Carini et al.\ (1991) intra-night
fluctuations in the atmospheric seeing could result in appreciably variable
light contributions from the host galaxy within the aperture.  Recently,
Cellone, Romero \& Combi (2000) argued that such spurious variations can be
substantial for AGN with bright galaxy hosts, particularly when small
photometric apertures are used.  Our DLCs are very unlikely to be affected by
this, since not only have we used sufficiently large apertures for photometry
(Table 2), but also all the QSOs in our sample are an order-of-magnitude more
luminous than their putative host galaxies, with the sole exception of the 
nearby LDQ 2349$-$014. In this case, host galaxy is seen on our CCD images;
hence, we used a rather large aperture ($6^{\prime \prime}$ radius). Moreover, we find
that either the seeing disk (as estimated from a star on the CCD frames,
which was thus monitored concurrently with the QSO), remained steady, or 
sharpened slightly during the night, implying that the observed gradual 
fading of the QSO (Figs.\ 1 and \ 2) cannot be understood in terms of any 
seeing variations during the nights the QSO was monitored.

Below we give brief remarks on individual QSOs for which we have positive or
probable detection of INOV, in increasing order of redshift.

\noindent{\bf LDQ 2349$-$014}, $z = 0.174$:  On each of the three nights 
the comparison stars S2 and S3 remained steady, and the QSO DLCs relative to 
both stars show a correlated decline by about 2\% over the roughly seven hours 
of monitoring (Fig.\ 1). 
Moreover, between 21.3 UT on 17 October 2001  and 13.5 UT on the next night 
the QSO brightened by $\sim$5\% (note the different ordinate labels).

\noindent{\bf LDQ 1004$+$130}, $z = 0.240$: This QSO was monitored on six
nights.
In Fig.\ 3 we present the DLCs for 24 March 2001, which shows a large spike 
of over 3\% ($\sim 9 \sigma$) at 16.23 UT relative to both available 
comparison stars, which themselves remained steady. The QSO brightened 
by $\sim$2\% between 20 UT on 29 March 2000 and the next night's 
observations which began at 16.0 UT (Stalin 2002; Stalin et al. 2003).

\begin{figure*}
\vspace*{-1.5cm}
\hspace*{-0.5cm}\psfig{file=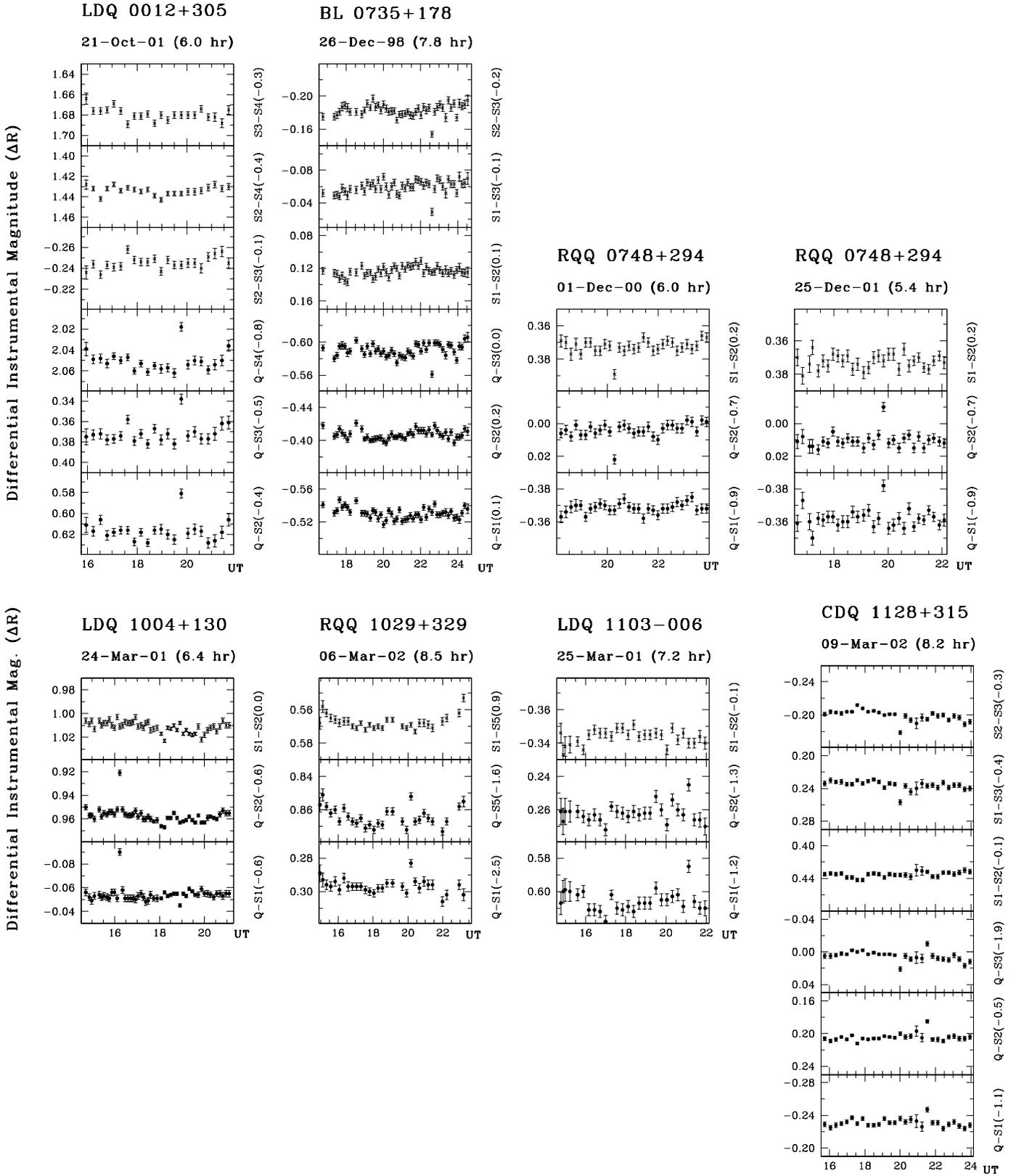,width=19.0cm}
\vspace*{-3.5cm}
\caption{DLCs for the dates on which single-point `spikes' were seen either 
on the QSO DLC or on the star-star DLC, displayed as in Fig.\ 1.}
\end{figure*}

\noindent{\bf RQQ 1252$+$020}, $z = 0.345$: Over the five nights of monitoring 
correlated variability of $\sim 2\%$ amplitude occured on 22 March 1999 
(Fig.\ 2). On 3 April 2000, the DLCs of the QSO against all three stars 
showed a gradual fading by $\sim$ 1\% during the 4.1 hours of monitoring 
(Fig.\ 2).  As discussed in GSSW03, this variation is well above the noise, 
and is opposite to that expected from the steady improvement observed
in the atmospheric seeing over the night.

\noindent{\bf RQQ 1101$+$319}, $z = 0.440$: Of the four nights this QSO 
was monitored, INOV was detected on 21 April 2001.  Against all three 
comparison stars, the QSO flux declined by about 1.5\% over the course of 
the night (Fig.\ 2). Although noisier than usual, the star-star DLCs do
not show any such systematic trend.

\noindent{\bf LDQ 1103$-$006}, $z = 0.426$: This QSO was observed on six 
nights. Unfortunately, only two reasonably steady comparison stars could be 
found. 
The DLCs of 18 March 1999 show probable INOV, where the LDQ dimmed 
by about 2\% during the first three hours of observations and thereafter
remained fairly steady (Fig.\ 1). On 6 April 2000 there was a probable 
variation involving an increase in QSO brightness of $\sim$ 1\% during 
the $\sim$ 4 hours of monitoring (Fig.\ 1). On 22 March 2002, this QSO 
showed a probable
variation (Fig.\ 1). On 25 March 2001, an apparent variability in the form 
of a spike of about 2\% ($\sim4 \sigma$) was seen at 21.13 UT (Fig.\ 3).

\noindent{\bf RQQ 1029$+$329}  (CSO 50), $z = 0.560$:  This QSO was observed 
on five nights, and DLCs for two of them are shown in Fig.\ 2.  On 5 
April 2000, the QSO slowly dimmed by about 2\% against all three stars during 
the course of the night; the seeing steadily improved during the night and, 
if of importance, would have yielded a small gradient opposite to that 
observed (GSSW03).  On 8 March 2002, against all three comparison stars
(in particular, S4 and S6 which were very steady), the RQQ showed a 
significant variability (Fig.\ 2), while the atmospheric seeing remained 
very steady throughout the session. 
During the remaining two nights (23 March 2001 and 6 March 2002) 
spikes of $\sim$1.7\% ($\sim$3--4$\sigma$) were detected (Fig. \ 3; Stalin et al. 
2003).

\noindent{\bf LDQ 0709$+$370}, $z = 0.487$: On one of the five nights this 
QSO was monitored, 20 December 2001, 
the QSO brightened by about 1\% against all three comparison stars during the 
first half of the night and then returned to its initial level during the 
second half of the night (Fig.\ 1).

\noindent{\bf RQQ 0748$+$294}, $z = 0.910$: This object was also observed 
by us earlier, and probable (spike) INOV was noted (Paper IV).  During the
present campaign, it was monitored for six nights, of which the last four had 
excellent sensitivity and coverage.
On 25 December 2001 the QSO DLCs showed a $\sim$ 2\% spike at 19.83 UT 
($\sim 4\sigma$) (Fig.\ 3).  Also, a significant star spike was observed at 
20.28 UT on 1 December 2000 (star 2 rose by $\sim$1.8\%; 4$\sigma$) (Fig.\ 3).

\noindent{\bf LDQ 0012$+$305}, $z = 1.619$: All measurements over six nights 
were fairly noisy due to the faintness of the QSO ($m_B = 17.2$).
A spike of 4\% ($> 4\sigma$) was seen for this QSO at 19.78 UT on 21 October 
2001 (Fig.\ 3).

\subsection{Long term optical variability (LTOV)}

Here we comment upon the subset of the four RQQs and the three LDQs for
which significant changes were found in their R-band flux over the longer 
period we monitored them. In increasing order of redshift they are:

\noindent{\bf RQQ 0945$+$438, $z = 0.226$}: Between 26 February 2000 and 
23 January 2001 this QSO was found to have dimmed by about 0.07 mag.

\noindent{\bf LDQ 1004$+$130, $z = 0.240$}:  A drop of 0.09 mag is observed 
between 16 March 1999 and 29 March 2000.  No level fluctuations exceeding
0.02 mag were noticed in four subsequent epochs of observations (Fig.\ 4).

\noindent{\bf RQQ 1252$+$020, $z = 0.345$}:  This QSO was observed on 5 nights
over a three year period beginning 22 March 1999.  It  brightened by
0.18 mag between 3 April 2000 and 26 April 2001, and had faded by 0.10 mag 
at the time of the last observation on 18 March 2002.

\noindent{\bf LDQ 1103$-$006, $z = 0.426$}: This QSO was monitored on 6 nights 
and was steady for the first three of them, from 17 March 1999 to 
6 April 2000 (Fig.\ 5).  By the time of the next observation on 25 March 2001,
it had brightened by 0.32 mag, and both subsequent measurements (until 
22 March 2002) found it at the same level (to within 0.25\%).

\noindent{\bf RQQ 1101$+$319, $z = 0.440$}: A drop of 0.2 mag over the course of
roughly one year (12 March 1999 to 4 April 2000) was seen, with a rise of 
about 0.07 mag observed by the time of the next observation on 21 April 2001.

\noindent{\bf RQQ 0748$+$294, $z = 0.910$}: We monitored this RQQ on 6 nights 
between 14 December 1998 and 25 December 2001 (Fig.\ 6). The comparison 
stars were always stable to within 1\%.  A dip of 0.23 mag was found 
between 14 December 1998 and 13 January 1999; by the time of the next 
observations on 9 December 1999 the source had recovered to its original 
brightness level, and was found at the same level (to within 0.25\%) in our 
subsequent measurements during the following two years (Sect.\ 5.3).

\noindent{\bf LDQ 0012$+$305, $z = 1.619$}:  This LDQ was monitored on three 
nights in January 2001 and on another three nights in October 2001. It remained 
stable within each month, but dropped by 7\% during the intervening period 
(between January 2001 and October 2001).

\begin{figure}
\vspace*{-5.5cm}
\hspace*{-1.0cm}\psfig{file=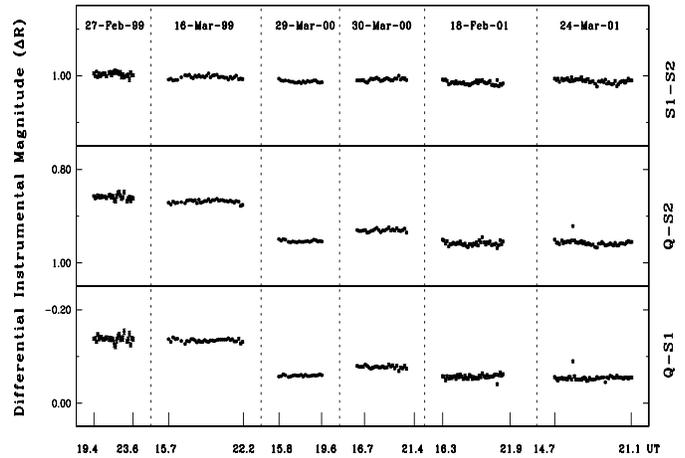,height=14cm,width=11cm}
\vspace*{-2.0cm}
\caption{
Long-term variations in the DLCs of the LDQ 1004$+$130, with all star-star
and QSO-star DLCs plotted as labelled.}
\end{figure}

\begin{figure}
\vspace*{-6.0cm}
\hspace*{-1.0cm}\psfig{file=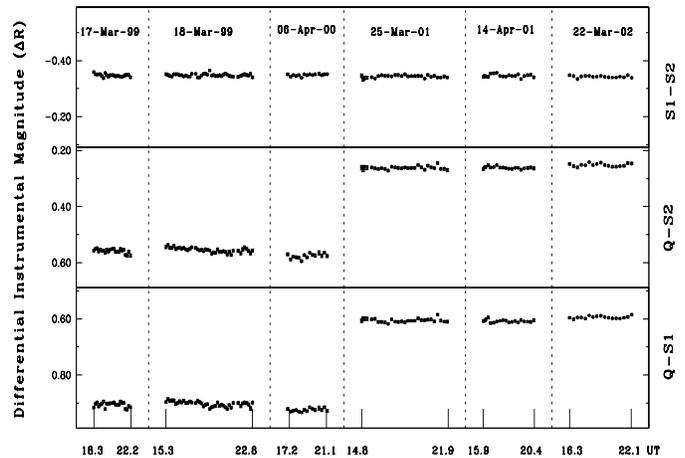,height=14cm,width=11cm}
\vspace*{-2.0cm}
\caption{
Long-term variations in the DLCs of the LDQ 1103$-$006, displayed as in
Fig.~4.}
\end{figure}

\begin{figure}
\hspace*{-1.0cm}\psfig{file=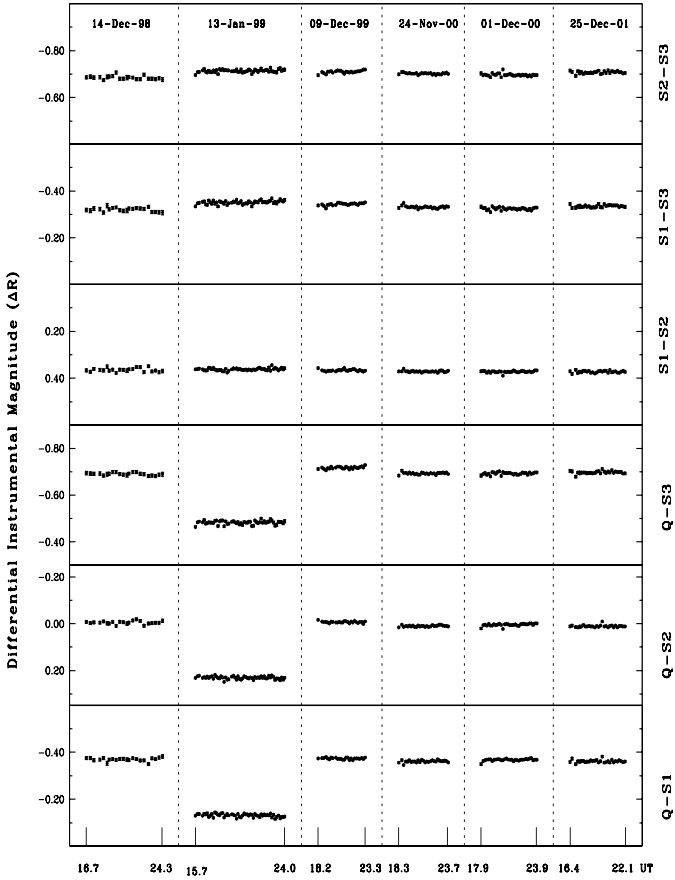,height=14cm,width=11cm}
\vspace*{-1.5cm}
\caption{
 Long-term variations in the DLCs of the  RQQ 0748$+$294, displayed as in
Fig.~4.}
\end{figure}

\section{Discussion}

\subsection{Intra-night Optical Variability (INOV)}

In Sect.\ 2 we mentioned the marked diversity in the estimates of duty cycles for
INOV in various classes of AGN.  Romero et al.\ (1999) estimated duty cycles 
for RQQs of only about 3\%, while for X-ray selected blazars this duty cycle 
was 28\% and for radio selected quasars (including blazars) 72\%, which is 
comparable to the estimates (over 80\%) by Carini (1990) and Heidt \& Wagner 
(1996) for (radio loud) blazars.  However, de Diego et al.\ (1998) concluded
that the probabilities of detecting INOV for RQQs and CDQs were statistically 
indistinguishable;  their implied value for our typical 6 hour duration of 
monitoring would be $\sim$25\%.

Both classes of objects in our matched sample were monitored with equally 
high sensitivities for comparable periods: the mean observing time for RQQs 
was $6.4 \pm 1.2$ hours per night over 29 nights, while for LDQs it was 
$5.9 \pm 1.5$ hours per night over 32 nights. Thus it is possible to directly 
compare the INOV duty cycles for these two AGN classes.  Following Romero et 
al.\ (1999) we define the duty cycle, DC, so that the contribution to the duty
cycle has been weighted by the number of hours (in its rest frame) for which 
each source was monitored,
\begin{equation}
DC = 100 \frac{\sum_{i=1}^{n} N_i(1/\Delta t_i)}{\sum_{i=1}^{n}(1/\Delta t_i)}
\ \     \%,
\end{equation}
\noindent where
$\Delta t_i = \Delta t_{i,obs}(1 + z)^{-1}$ is the duration (corrected for 
cosmological redshift) of an $i_{th}$ monitoring session of the source out of 
a total of $n$ sessions for the selected AGN class, and $N_i$ equals 0 or 1, 
depending on whether the object was  non-variable or 
variable, respectively, during $\Delta t_i$.

For RQQs, counting only observing sessions for which the INOV was clearly
detected, we find the DC = 17\%.  
Our value falls roughly mid-way between the lower estimates published by 
Jang \& Miller (1997) and by Romero et al.\ (1999), and the higher estimate 
of de Diego et al.\ (1998), though we note that the latter analysis technique 
is less trustworthy (Sect. \ 2).

Turning to LDQs, we find a DC of 9\% for a clear detection of INOV. An additional
possible contribution of 6\% comes from the two cases of probable detection. 
So the  DC including all likely INOV for the LDQs is about 15\%. Given the 
rather small number of detections that resulted despite the substantial length 
of our observations, one clearly cannot claim any statistically significant
difference between the DCs of INOV for the RQQ and LDQ classes. 
Interestingly, the ranges of INOV amplitudes for both RQQs and LDQs are
also found to be very similar ($\psi < 3\%)$ (Table 2; Sect.\ 4.1) and this is 
a key result of the present study. It is also noteworthy that a close similarity 
between LDQs and RQQs in terms of INOV duty cycle extends even to
BL Lac objects if only their small-amplitude INOV ($\psi < 3\%$) is 
considered (see, Fig. 2 of GSSW03). A possible explanation of these similarities in 
the INOV characteristics is outlined below.

A key motivation of our program was to assess the role of relativistic 
beaming in the INOV of AGN classes other than blazars for which the bulk of 
all variability is believed to arise from instabilities in their relativistic 
jets. In our monitoring program, both the INOV duty cycle and amplitudes for 
BL Lacs are found to be much higher than for the RQQs (GSSW03) and also
compared to LDQs (present work). Therefore it is relevant to ask to what 
extent the modest variations of the RQQs and LDQs might be understood 
within the conventional relativistic jet paradigm if one postulated that 
such jets exist (on optically emitting scale lengths) even in RQQs, and
also accepts the conventional wisdom that, in general, the axes of QSOs are 
mildly misaligned from us (e.g., Barthel 1989; Antonucci 1993). In GSSW03, 
we argued in favour of such a possibility and, likewise, we now explore this
point further, taking a clue from the BL Lac object OJ287 monitored by us.

For objects whose flux densities are relativistically beamed, the observed
degree of flux variability and consequently, the duty cycles, can be strongly
influenced by the beaming, since any intrinsic flux variations associated with 
the relativistic outflow will have their time-scales shortened and amplitudes 
boosted in the observer's frame.  As usual, the Doppler factor is defined 
as $\delta = [\Gamma(1 - \beta {\rm cos}\theta)]^{-1}$,
where $\beta = v/c$, $\Gamma = (1-\beta^2)^{-1/2}$ is the bulk Lorentz factor 
of the jet, and $\theta$ is the viewing angle. Then the observed flux density, 
$S_{obs}$ is given in terms of the intrinsic flux density, $S_{int}$ 
(e.g.\ Urry \& Padovani 1995)
\begin{equation}
S_{obs} = \left(\frac{\delta}{1 + z}\right)^p S_{int}. 
\end{equation}
Here, $p = 3-\alpha$ for a moving disturbance or blob in the jet; the spectral 
index of the emission, $\alpha \equiv  {d{\rm ln}(S_{\nu})/d{\rm ln}(\nu)}$
and we have assumed $\alpha = -1$ when evaluating this expression
(Stocke et al.\ 1992). Similarly, due to the beaming, the observed time scale 
becomes shortened as $\Delta t_{obs} = \Delta t_{int} (1 + z) / \delta$.

\begin{figure}
\vspace*{-2.0cm}
\hspace*{-1.0cm}\psfig{file=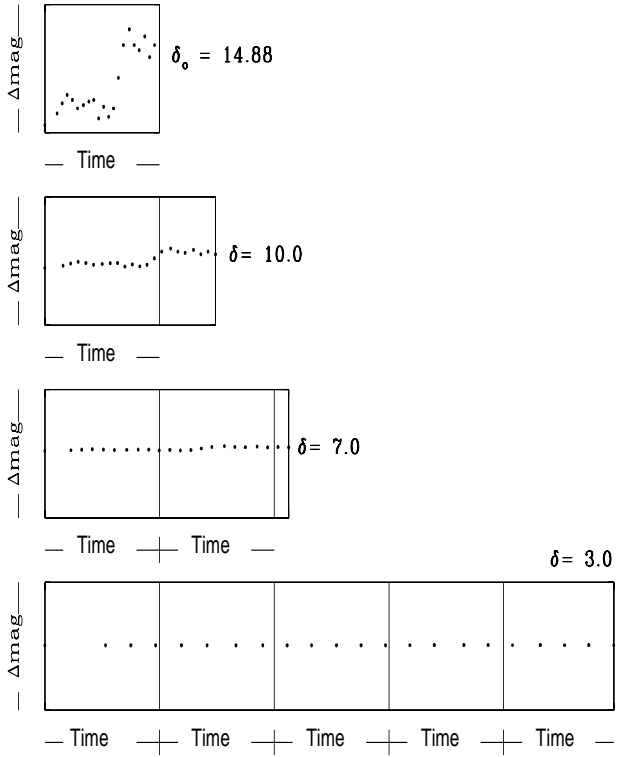,height=14cm,width=10cm}
\vspace*{-1.5cm}
\caption{
The top panel presents the R-band light curve of the BL Lac object 0851$+$202 
(OJ 287), showing a $\sim$ 5\% fluctuation,  observed on 28 March 2000.
Subsequent panels show its translation to progressively smaller Doppler 
factors, corresponding to larger viewing angles, assuming the model of a 
discrete moving blob in the relativistic jet. Each rectangular window shows 
an equal time interval (4.2 hours in this case) in the observer's frame, 
and the total amplitude range, $\Delta$ mag, for each panel is 0.1-mag., 
relative to the same comparison star.}
\end{figure}

In Fig.\ 7 we illustrate the effect of the Doppler beaming on the observed
DLCs assuming the spherical blob model; conclusions for the continuous jet 
model are similar (GSSW03).  We start with a DLC of the BL Lac object 0851$+$202 
(OJ 287) which exhibited a large ($\sim$5\%) and rapid ($\sim$0.8 hour) variation 
on 28 March 2000 (Stalin 2002; Sagar et al.\ 2003). For this source, $\delta_o 
= 14.88$ has been estimated by Zhang, Fan \& Cheng (2002).
We now use this DLC to simulate light curves for lower values of $\delta$,  as
would be seen by a observers at a larger viewing angles to the jet, by mapping 
the observed DLC onto the `amplitude--time' plane corresponding to chosen 
values of $\delta$.  This mapping is achieved simply by compressing the 
observed DLC amplitudes by a factor $(\delta/\delta_o)^{p}$ and, 
simultaneously, stretching the DLCs along the time axis by a factor 
$(\delta_o/\delta)$, where $\delta_o$ and $\delta$ are the original (actual) 
and lower adopted values of the Doppler factor, respectively.

From these (partially) Doppler de-beamed DLCs (Fig.\ 7), it is evident that 
even an observer at only a marginally misaligned  direction to the jet will monitor 
a drastic reduction in both the amplitude and rapidity of the INOV for the 
same BL Lac object which appears highly variable to a better aligned 
observer.  For example, if $\theta = 3.^{\circ}75$ for the jet of OJ 287, 
the estimated $\delta_o = 14.88$ corresponds to $\beta = 0.9966$.
This would give $\delta = 10$ for a modestly misaligned jet
with $\theta \simeq 6^{\circ}$, $\delta = 7$ for  $\theta \simeq 7.^{\circ}5$ 
and $\delta = 3$ for  $\theta \simeq 13^{\circ}$; such misalignments (or even
somewhat larger ones corresponding to even smaller $\delta$) are believed to 
be typical of LDQs (Barthel 1989) and RQQs as well (Antonucci 1993). 
The simulated light curves for  viewing angles greater than $\sim 10^{\circ}$ 
show barely detectable INOV, as is indeed observed for both LDQs and RQQs 
(Figs. 1 \& 2).  From this we surmise that the mere absence of pronounced 
INOV in RQQs in no way rules out the possibility that they have optical 
synchrotron jets as active intrinsically (albeit somewhat more misdirected)
as those in BL Lacs (GSSW03). An independent support to this assertion comes 
from the similarity found here in the INOV of RQQs and LDQs,
since the central engines of LDQs are in any case believed to emit relativistic
synchrotron jets (e.g., Urry \& Padovani 1995).

\subsection{Spikes}

Strong single-point fluctuations, or spikes, were noted for several RQQs in 
Paper IV, and similar events were noted in our present monitoring campaign of 
AGN (Fig.\ 3).  We did not see such excursions 
in any of the star$-$star DLCs from the programme reported
in Paper IV. This led us to conclude that these events were
probably intrinsic to the RQQs, though we noted that simultaneous detection
at different sites was essential to confirm the reality of such events.  
However, our present measurements have
greater sensitivity, thanks to the new Wright CCD, so we were able
to make a careful search for this type of fluctuations in the entire
data set, including the DLCs of CDQs and BL Lacs that are reported
in detail elsewhere (Stalin et al.\ 2003; Stalin 2002).  In performing this
search we conservatively define a spike as a single point fluctuation
visible simultaneously in multiple DLCs involving the same object, after which 
the flux returns to essentially the pre-spike level and the amplitude of the 
fluctuation is a minimum of $2.5\sigma$ (corrected by the factor of 1.5, Sect. 
4.1) on at least two of the DLCs involving the object showing the spike.

Based on these criteria we have identified 15 spikes associated with the
AGNs in our sample (including all the four types) and 20 spikes associated with
their comparison stars.  Since we typically derive DLCs for 2 or 3 comparison
stars for each QSO, if these spikes were non-intrinsic random events, one 
would naively expect a couple of times as many spikes to be seen for the stars 
than for the QSOs.  But, since on average, the quasars are somewhat fainter
than the comparison stars, the relative number of detectable spikes would be
slightly enhanced for the QSOs.  We note that all but one of the spikes
was positive; the exception was for the Star 3 in the field of the 
LDQ 0709$+$370 on 20 January 2001 at 18.19 UT where a negative deviation
of $\sim$1.7\% ($> 5\sigma$) was found.

A single stellar spike in Star 2 for RQQ 0748$+$294, plotted in Fig.\ 3,
was mentioned above.  In Fig.\ 3 we also plot 
an additional two stellar spikes discovered in this wider data base, one 
from Star 3 in the comparison group for the BL Lac 0735$+$178 on 26 December 
1998 at 22.62 UT, and one from Star 3 in the comparison group of the 
CDQ 1128$+$315 on 9 March 2002 at 20.00 UT; note that this blazar also shows 
a spike at 21.53 UT on the same night.  A table containing the details of 
these spike data will appear in Stalin et al.\ (2003).

We have determined the flux density of each spike, using the R magnitudes of 
the corresponding star or AGN, as follows.  We convert the R magnitudes of 
the stars, taken from the USNO catalog, into flux densities and multiply by 
the average magnitude fluctuation of the spike.  For the spikes on the AGN, 
since the R magnitudes of the AGN can differ significantly from those tabulated 
in the USNO catalog (due to long-term variability), we determined these instead 
by adding our observed mean differential magnitudes (QSO$-$star) on the night of the 
spike to the USNO R-magnitudes of the corresponding stars and then taking an 
average of these values.

\begin{figure}
\vspace*{-0.5cm}
\hspace*{-1.0cm}\psfig{file=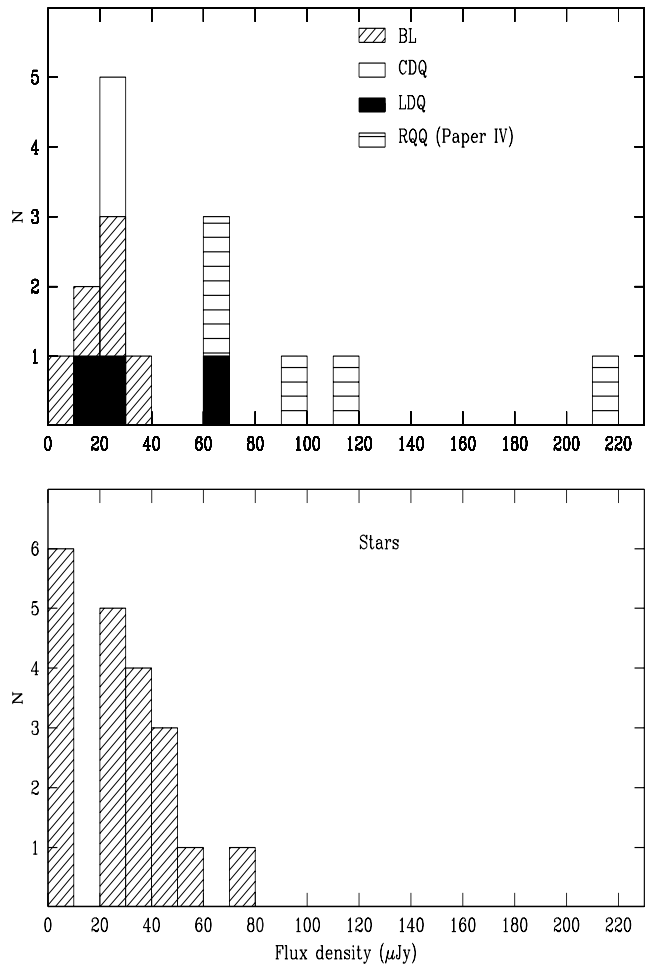,height=14cm,width=10cm}
\vspace*{-0.5cm}
\caption{
Histogram of flux densities of the ``spikes'' detected in both the current 
programme and those reported in Paper IV.  The lower panel gives the 
distribution of fluxes for the spikes seen in the DLCs between the comparison 
stars and the upper panel shows the same as seen in the DLCs between the 
different classes of AGN and the comparison stars.} 
\end{figure}

Fig.\ 8 shows a histogram of the flux densities attributable to these
various spikes.  In that our more sensitive observations have detected many 
cases of spikes even in the DLCs of the comparison stars, our earlier tentative 
conclusion (Paper IV) that since these had been found to occur only in the RQQs, 
and were therefore likely to be intrinsic to the RQQs, is not confirmed. 
If, as indicated by Fig.\ 8,  these spikes are practically as common in QSOs as 
in stars, the simplest interpretation would be that they are caused by 
compact cosmic ray hits.  It is also possible that some of them are of 
unknown instrumental origin. Yet it is worth noting that six of the seven 
most powerful spikes (those above 60$\mu$Jy) are associated with the QSOs 
and not with the stars, so that we may still be observing some intrinsic
ultra-rapid QSO variability.  Possible explanations for such extreme events 
include brief periods during which coherent emission processes can be important
(e.g.\ Lesch \& Pohl 1992; Krishan \& Wiita 1990, 1994; Kawaguchi et al.\ 1998) 
or perhaps extreme turbulence and fortuitous intense Doppler boosting of emission 
from a dense or highly magnetized portion of the jet flow temporarily moving very 
close to the line of sight (e.g.\ Gopal-Krishna \& Wiita 1992).

\subsection{Long-term optical variability (LTOV)}

As expected (e.g.\ Smith et al.\ 1990) nearly all QSOs which we monitored for
extensive periods ($> 2$ months) exhibited significant amounts of LTOV.
The RQQs 0945$+$438, 1252$+$020, 0748$+$294 and 1101$+$309 as well as the
LDQs 1004$+$130, 1103$-$006, and 0012$+$305 exhibited significant changes, 
as described in Sect.\ 4.2.  In addition, the LDQ 0709$+$370 showed a 
variation of about 3\% between January and December 2001.

The only object not observed to have varied over month/year timescales is the 
RQQ 1029$+$329 (CSO 50), which was observed on five nights over the span of
two years; it did however, show weak but clear INOV on two of these nights
(GSSW03; Fig.\ 2).  
The other five objects in our sample (RQQs: 0514$-$005, 1017$+$279; LDQs: 
2349$-$014, 0134$+$329, 0350$-$073) were only observed for total time 
baselines between 4 days and 6 weeks, so the lack of LTOV detection for 
them is not surprising.

Although our long-term sampling was not frequent enough to allow estimates 
for possible physical timescales to be obtained, we must remark upon the 
peculiar behaviour of the RQQ 0748$+$294 ($z = 0.910$) which was rather
evenly smapled between December 1998 and December 2001 (Fig.\ 6). For the first 
and the last four measurements this object showed a constant brightness to 
within one percent, but on the night of 13 January 1999 it was about 0.2 mag 
fainter.  While a relatively brief brightening of this order is explicable 
in many scenarios, such a dip is somewhat unexpected.  We suggest that this
dimming may correspond to a partial eclipse of the optical continuum source 
with a duration of less than six months in the source frame.  A sufficiently 
warped or otherwise thickened portion of the accretion disk at roughly 100 
Schwarzschild radii from a $10^8$M$_\odot$ putative central black hole could  
block a significant fraction of the optical continuum emission from the inner 
portion of the disk for some months.

While our sample is rather small, there is no obvious difference in the 
behaviour of RQQs and LDQs in terms of optical variability on month-to-year 
timescales.  This is in accord with the conclusions reached by Paltani \& 
Courvoisier (1994) from their study of IUE data for radio-loud and radio-quiet 
QSOs in the ultraviolet.

\section{Conclusions}

The observations reported here in detail have placed on a firm footing
the phenomenon of optical intra-night variability of optically luminous 
quasars of non-blazar type (both radio-quiet and lobe-dominated; also see 
GSSW03). The dense temporal sampling over long duration, together with the 
good sensitivity attained in our campaign using CCDs as N-star photometers, 
has clearly demonstrated that small amplitude (typically 0.01--0.03 mag) 
variations on time-scales of hours are real. Even though the percentage 
luminosity variation implied by the INOV of these luminous RQQs and LDQs 
is small, the total power involved is still so enormous as to render a 
starburst/supernova explanation (e.g.\ cid Fernandes et al.\ 1996) untenable 
for these rapid events.

Although our full programme covered BL Lacs and CDQs as well (see, GSSW03;
Sagar et al.\ 2003; Stalin et al.\ 2003), in this paper we have provided 
results for a 
sub-sample of 7 RQQs and 7 LDQs matched in both redshift and optical power.
A key result of our observations is that there is no significant difference 
in either the amplitude or duty-cycle of INOV between these two classes of 
non-blazar AGN. We thus infer that {\it the radio loudness of a quasar alone
is not a sufficient condition for a pronounced INOV}. 

Secondly, our expanded study of single-point fluctuations which occurred on
the time scale smaller than $\sim$15 min (including those seen on the DLCs 
of  CDQs and BL Lacs) leads to the conclusion that, with the possible 
exception of the strongest ones, most of the spikes are probably caused by 
cosmic-ray hits.  To confirm the origin of such spikes, it would be desirable 
to try to simultaneously observe such events from two independent observatories.

We have also presented some limited, but interesting, results on longer-term 
(month to year) variability of these non-blazar AGNs.  Again we find no 
significant difference between the RQQs and LDQs, for both of which long-term 
variability is similarly common.  We speculate that the 0.2 mag dip in the 
light-curve of the RQQ 0748$+$294 on 13 January 1999 is due to partial 
occultation of the optical continuum source, perhaps by a non-axisymmetric 
disk deformation.

While our data cannot exclude accretion disk flares as the source of the much 
milder and rarer intranight optical variability observed for RQQs and LDQs,
as compared to BL Lacs (Sect.\ 1), it does not preclude a substantial 
contribution to this type of flux variability coming from blazar-like 
relativistically beamed emission.  In Sect. 5.1 and GSSW03 we have demonstrated 
that a typical RQQ light-curve can be derived from an observed blazar 
light-curve even if the RQQ possesses a jet intrinsically as active as a BL Lac
jet, albeit observed at a modest offset in the viewing angle ($\sim 10^{\circ} - 
20^{\circ}$). Such a scenario would also be consistent with the similarity
found here between the INOV of RQQs and LDQs, since the latter are already
believed to have central engines ejecting relativistic synchrotron jets.
Inverse Compton quenching of the jets in a majority of quasars before reaching
the scale probed by radio emission (e.g.\ Brown 1990) could, conceivably,
be responsible for the large difference between the radio luminosties of
radio-loud and radio-quiet quasars (GSSW03).
A possible signature of such quenching is the hard X-ray spectral tail found 
in some RQQs (George et al. 2000). This emission is seen despite the extremely 
strong forward 
flux boosting of the X-rays expected from the inverse Compton scattering of 
external (e.g. BLR) photons by the relativistic jet 
($\propto \delta^{(4 - 2\alpha)}$, Dermer 1995). In this fashion, radio-loud 
and radio-quiet quasars can be unified through an orientation based scheme, at 
least in the realm of intra-night optical variability. This picture is
 broadly in accord with the idea of {\it jet-disk symbiosis}, where jets
of some type are to be expected from essentially any type of disk 
(e.g., Falcke, Malkan \& Biermann 1995).

\section{Acknowledgments} We thank John McFarland for a helpful discussion.
This research has made use of the NASA/IPAC Extragalactic Database (NED),
which is operated by the Jet Propulsion Laboratory, California Institute 
of Technology, under contract with the National Aeronautics and Space 
Administration. G-K is appreciative of hospitality at GSU and is also grateful 
for support from the ESO senior visitor program during which this programme 
originated a decade ago. CSS thanks NCRA for hospitality and the use of its 
facilities.  PJW is grateful for hospitality at NCRA and at the Princeton 
University Observatory. PJW's efforts were partially supported by Research 
Program Enhancement funds at GSU.

\end{document}